\documentclass[aps,prl,showpacs,showkeys,
twocolumn,superscriptaddress,altaffilletter,
 amsmath,amssymb]{revtex4-1}
\usepackage{graphicx}
\usepackage{dcolumn}
\usepackage{bm}
\usepackage{upgreek}
\usepackage{hyperref}
\usepackage{natbib}
\usepackage[mathlines]{lineno}
\newcommand{\ctsper}      {cts/(keV$\cdot$kg$\cdot$yr)}

\newcommand{\dctsper}     {{$10^{-3}$~cts/(keV$\cdot$kg$\cdot$yr)}}

\newcommand{\kgyr}        {{kg$\cdot$yr}}

\newcommand{\mus}         {{$\upmu$s}}

\newcommand{\qbb}         {{$Q_{\beta\beta}$}}
\newcommand{\thalfzero}   {${T^{0\nu}_{1/2}}$}

\newcommand{\onbb}        {{$0\nu\beta\beta$}}

\newcommand{\twonu}       {{$2\nu\beta\beta$}}

\newcommand{\fgesix}      {\mbox{$f_{76}$}}

\newcommand{\factvol}     {\mbox{$f_{av}$}}

\newcommand{\gerda}       {\textsc{Gerda}}

\newcommand{\GERDA}       {\mbox{\textsc{Gerda}}}  
\newcommand{\igex}        {\textsc{Igex}}
\newcommand{\IGEX}        {{\mbox{\textsc{Igex}}}}
\newcommand{\hdm}         {\textsc{HdM}}

\newcommand{\gesix}       {{$^{76}$Ge}}
\newcommand{\gess}        {{$^{76}$Ge}}
\newcommand{\geenr}       {{$^{\rm enr}$Ge}}          

\newcommand{\exposure}    {\mbox{$\cal E$}}

\newcommand{\effpsd}      {\mbox{$\varepsilon_{psd}$}}

\begin{document}
\title{
Results on neutrinoless double beta
decay of $^{76}$Ge from {\sc Gerda} Phase~I
}
%
%
  \affiliation{INFN Laboratori Nazionali del Gran Sasso, LNGS, Assergi, Italy}
  \affiliation{Institute of Physics, Jagiellonian University, Cracow, Poland}
  \affiliation{Institut f{\"u}r Kern- und Teilchenphysik, Technische Universit{\"a}t Dresden, Dresden, Germany}
  \affiliation{Joint Institute for Nuclear Research, Dubna, Russia}
  \affiliation{Institute for Reference Materials and Measurements, Geel, Belgium}
  \affiliation{Max-Planck-Institut f{\"u}r Kernphysik, Heidelberg, Germany}
  \affiliation{Dipartimento di Fisica, Universit{\`a} Milano Bicocca, Milano, Italy}
  \affiliation{INFN Milano Bicocca, Milano, Italy}
  \affiliation{Dipartimento di Fisica, Universit{\`a} degli Studi di Milano e INFN Milano, Milano, Italy}
  \affiliation{Institute for Nuclear Research of the Russian Academy of Sciences, Moscow, Russia}
  \affiliation{Institute for Theoretical and Experimental Physics, Moscow, Russia}
  \affiliation{National Research Centre ``Kurchatov Institute'', Moscow, Russia}
  \affiliation{Max-Planck-Institut f{\"ur} Physik, M{\"u}nchen, Germany}
  \affiliation{Physik Department and Excellence Cluster Universe, Technische  Universit{\"a}t M{\"u}nchen, Germany}
  \affiliation{Dipartimento di Fisica e Astronomia dell'Universit{\`a} di Padova, Padova, Italy}
  \affiliation{INFN Padova, Padova, Italy}
  \affiliation{Physikalisches Institut, Eberhard Karls Universit{\"a}t T{\"u}bingen, T{\"u}bingen, Germany}
  \affiliation{Physik Institut der Universit{\"a}t Z{\"u}rich, Z{\"u}rich, Switzerland}

\author{M.~Agostini}
  \affiliation{Physik Department and Excellence Cluster Universe, Technische  Universit{\"a}t M{\"u}nchen, Germany}
\author{M.~Allardt}
  \affiliation{Institut f{\"u}r Kern- und Teilchenphysik, Technische Universit{\"a}t Dresden, Dresden, Germany}
\author{E.~Andreotti}
  \affiliation{Physikalisches Institut, Eberhard Karls Universit{\"a}t T{\"u}bingen, T{\"u}bingen, Germany}
  \affiliation{Institute for Reference Materials and Measurements, Geel, Belgium}
\author{A.M.~Bakalyarov}
  \affiliation{National Research Centre ``Kurchatov Institute'', Moscow, Russia}
\author{M.~Balata}
  \affiliation{INFN Laboratori Nazionali del Gran Sasso, LNGS, Assergi, Italy}
\author{I.~Barabanov}
  \affiliation{Institute for Nuclear Research of the Russian Academy of Sciences, Moscow, Russia}
\author{M.~Barnab\'e Heider}
\altaffiliation[presently at: ]{CEGEP St-Hyacinthe, 
 Qu{\'e}bec, Canada}
  \affiliation{Max-Planck-Institut f{\"u}r Kernphysik, Heidelberg, Germany}
  \affiliation{Physik Department and Excellence Cluster Universe, Technische
    Universit{\"a}t M{\"u}nchen, Germany}
\author{N.~Barros}
  \affiliation{Institut f{\"u}r Kern- und Teilchenphysik, Technische Universit{\"a}t Dresden, Dresden, Germany}
\author{L.~Baudis}
  \affiliation{Physik Institut der Universit{\"a}t Z{\"u}rich, Z{\"u}rich, Switzerland}
\author{C.~Bauer}
  \affiliation{Max-Planck-Institut f{\"u}r Kernphysik, Heidelberg, Germany}
% ...

\collaboration{{\textsc{Gerda} collaboration}}
\email{Correspondence: gerda-eb@mpi-hd.mpg.de}
\noaffiliation

\date{\today}
\begin{abstract}
Neutrinoless double beta decay is a process that violates lepton number
conservation. It is predicted to occur in extensions of the Standard Model of
particle physics.  This Letter reports the results from Phase~I of the
GERmanium Detector Array (\gerda) experiment at the Gran Sasso Laboratory
(Italy) searching for neutrinoless double beta decay of the isotope
\gess. Data considered in the present analysis have been collected between
November 2011 and May 2013 with a total exposure of 21.6~\kgyr.  A blind
analysis is performed.  The background index is about $1\cdot
10^{-2}$~\ctsper\ after pulse shape discrimination.  No signal is observed and
a lower limit is derived for the half-life of neutrinoless double beta decay
of \gess, \thalfzero$> 2.1\cdot 10^{25}$~yr (90~\% C.L.).  The combination
with the results from the previous experiments with \gess\ yields \thalfzero$
> 3.0\cdot 10^{25}$~yr (90~\% C.L.).
\end{abstract}

\vfill

\maketitle
\section{Introduction}

For several isotopes beta decay is energetically forbidden but the
simultaneous occurrence of two beta decays (\twonu) is allowed.  This process
has been observed in eleven nuclei with half-lives in the range of $10^{18} -
10^{24}$~yr~\cite{bar10,tre02}.  Extensions of the Standard Model predict that
also neutrinoless double beta (\onbb) decay should exist:
(A,Z)$\rightarrow$(A,Z+2)$+2 e^-$.  In this process lepton number is violated
by two units and the observation would have far-reaching
consequences~\cite{bil12,ver12,rodejohann,gomez12}.  It would prove that
neutrinos have a Majorana mass component.  Assuming the exchange of light
Majorana neutrinos, an effective neutrino mass can be evaluated by using
predictions for the nuclear matrix element (NME).

The experimental signature of \onbb\ decay is a peak at the $Q$-value of the
decay.  The two most sensitive experiments with the candidate nucleus
\gess\ ($Q_{\beta\beta}=2039.061\pm0.007$~keV~\cite{qbb}) were
Heidelberg-Moscow (\hdm)~\cite{hdm} and the International GErmanium eXperiment
(\igex)~\cite{igex,igex2}.  They found no evidence for the \onbb\ decay of
\gess\ and set lower limits on the half-life \thalfzero\ $>1.9\cdot10^{25}$~yr
and $>1.6\cdot10^{25}$~yr at 90\,\% C.L., respectively. Part of
\hdm\ published a claim to have observed $(28.75\pm6.86)$
\onbb\ decays~\cite{klapdor1} and reported \thalfzero$=(1.19^{+0.37}_{-0.23})
\cdot 10^{25}$~yr. Later, pulse shape information was used to strengthen the
claim~\cite{klapdor2}.  Because of inconsistencies in the latter reference
pointed out recently~\cite{bernhard}, the present comparison is restricted to
the result of Ref.~\cite{klapdor1}.

Until recently, the claim has not been scrutinized.  The currently most
sensitive experiments are KamLAND-Zen~\cite{kamland} and EXO-200~\cite{exo}
looking for \onbb\ decay of $^{136}$Xe and \GERDA~\cite{gerda:2013:tec}
employing \gess. Nuclear matrix element calculations are needed to relate the
different isotopes.  Thus the experiments using $^{136}$Xe can not refute the
claim in a model-independent way. \GERDA\ is able to perform a direct test
using the same isotope and also using mostly the same detectors as \hdm\ and
\igex.  This paper reports the \onbb\ results of Phase~I of \GERDA.

\section{The experiment}

The \GERDA\ experiment~\cite{gerda:2013:tec} is located at the Laboratori
Nazionali del Gran Sasso (LNGS) of INFN in Italy.  High-purity germanium
(HPGe) detectors made from isotopically modified material with
\gesix\ enriched to $\sim$86\,\% (\geenr) are mounted in low-mass copper
supports and immersed in a 64~m$^3$ cryostat filled with liquid argon (LAr).
The LAr serves as cooling medium and shield against external backgrounds. The
shielding is complemented by 3~m of water which is instrumented with photo
multipliers to detect Cherenkov light generated by muons.  The HPGe detector
signals are read out with custom-made charge sensitive amplifiers optimized
for low radioactivity which are operated close to the detectors in LAr.  The
analog signals are digitized with 100~MHz Flash ADCs and analyzed offline.  If
one of the detectors has an energy deposition above the trigger threshold
(40-100~keV), all channels are analyzed for possible coincidences.
 
Reprocessed $p$-type semi-coaxial detectors from the \hdm\ and
\IGEX\ experiments were operated together with newly produced \GERDA\ Phase~II
detectors.  The latter are of BEGe type manufactured by
Canberra~\cite{canberra}. The active volume fraction \factvol\ of the
detectors was determined beforehand amounting to 0.87 (0.92) for the
semi-coaxial (BEGe) detectors~\cite{gerda:2013:tec,gerda:2013:bck}.

Data acquisition started in November 2011 with eight \geenr\ detectors
(ANG~1-5 from \hdm\ and RG~1-3 from \igex), totaling a weight of 17.67~kg.
Five enriched \GERDA\ Phase~II detectors of 3.63~kg in total were deployed in
July 2012.  ANG~1 and RG~3 started to draw leakage current soon after their
deployment, and are omitted in this analysis.  One BEGe detector showed an
unstable behavior and is omitted as well.  Since March 2013, RG~2 is no longer
used since it is operated below its full depletion voltage.  A fraction of
5\,\% of the data was discarded because of temperature-related instabilities.
Results from the data collected until May 2013 (492.3~ live days) are reported
here.  The total exposure considered for the analysis amounts to
21.6~\kgyr\ of \geenr\ detector mass, yielding $(215.2\pm7.6)$~mol$\cdot$yr of
\gess\ within the active volume.

The offline analysis of the digitized charge pulses is performed with the
software tool \textsc{Gelatio}~\cite{gelatio} and the procedure described in
Ref.~\cite{acat}.  The deposited energy is reconstructed by a digital filter
with semi-Gaussian shaping.  Events generated by discharges or due to
electromagnetic noise are rejected by a set of quality cuts.

The energy scale of the individual detectors is determined with $^{228}$Th
sources once every one or two weeks.  The differences between the
reconstructed peak positions and the ones from the calibration curves are
smaller than 0.3~keV.  The energy resolution was stable over the entire data
acquisition period. The gain variation between consecutive calibrations is
less than 0.05\,\%~\cite{gerda:2013:tec}, which corresponds to $<30\,\%$ of
the expected energy resolution (Full Width Half Maximum, FWHM) at
\qbb. Between calibrations, the stability is monitored by regularly injecting
charge pulses into the input of the amplifiers.

The energy spectrum and its decomposition into individual sources is discussed
in Ref.~\cite{gerda:2013:bck}. Peaks from $^{40}$K, $^{42}$K, $^{214}$Bi,
$^{214}$Pb and $^{208}$Tl $\gamma$ rays can be identified as well as $\alpha$
decays from the $^{226}$Ra decay chain, and $\beta$ events from $^{39}$Ar. All
$\gamma$-ray peaks are reconstructed at the correct energy within their
statistical uncertainty.  The energy resolution (FWHM) of the strongest line
(1524.6~keV from $^{42}$K) is 4.5 (3.1)~keV for the semi-coaxial (BEGe)
detectors. These values are about 10\,\% larger than the resolutions obtained
from calibrations. The broadening is due to fluctuations of the energy scale
between calibrations.  The interpolated FWHM at \qbb\ for physics data is
detector dependent and varies between 4.2 and 5.7~keV for the semi-coaxial
detectors, and between 2.6 and 4.0~keV for the BEGe detectors. The
exposure-averaged values are $(4.8\pm0.2)$~keV and $(3.2 \pm 0.2)$~keV,
respectively.

For the first time in the field of \onbb\ decay search, a blind analysis was
performed in order to avoid bias in the event selection criteria.  Events with
energies within \qbb$\pm 20$~keV were not processed. After the energy
calibration and the background model were finalized the window was opened
except for $\pm5$~keV ($\pm4$~keV) around \qbb\ for the semi-coaxial (BEGe)
detectors. After all selections discussed below had been frozen, the data in
the \qbb\ region were analyzed.  The validity of the offline energy
reconstruction and of the event selection procedures have been cross-checked
with a fully independent analysis.

\section{\onbb\ analysis} 
The signature for \onbb\ decay is a single peak at \qbb. Furthermore, events
from \onbb\ decays have a distinct topology, which allows to distinguish them
from $\gamma$-induced background. For \onbb\ events, energy is deposited by
the two electrons, which have a short range in germanium: more than 90\,\% of
\onbb\ events are expected to deposit all energy localized within few mm$^3$
(single-site events, SSE).  On the other hand, most background events from
$\gamma$-ray interactions have energy depositions in many detectors or at
different, well separated, positions (multi-site events, MSE).

Only events with an energy deposition in a single detector are accepted
resulting in a background reduction by about 15\,\% around \qbb, with no
efficiency loss for \onbb\ decays. Events in the HPGe detectors are rejected
if they are in coincidence within 8~\mus\ with a signal from the muon
veto. This leads to a further background reduction by about 7\,\%.  Events
which are preceded or followed by another event in the same detector within
1~ms are excluded. This allows to reject background events from the
$^{214}$Bi-$^{214}$Po cascade (BiPo) in the $^{222}$Rn decay chain.  Less than
1\,\% of the events at \qbb\ are affected by this cut. Due to the low counting
rate in \gerda\ and due to the low muon flux at LNGS, the dead time due to the
muon veto and BiPo cuts is negligible.

The detector signals are different for SSE and MSE, and also surface events
from $\beta$ or $\alpha$ decays exhibit a characteristic shape. Thus, pulse
shape discrimination (PSD) techniques can improve the sensitivity.

For BEGe detectors, a simple and effective PSD is based on the ratio of the
maximum of the current pulse (called $A$) over the energy
$E$~\cite{aovere,Agostini:2010rd,gerda:2013:psd}.  The $A/E$ cut efficiency is
determined from calibration data using events in the double escape peak (DEP)
of the 2615~keV $\gamma$ ray from $^{208}$Tl. It is cross-checked with
\twonu\ decays of \gess.  The acceptance of signal events at \qbb\ is
\effpsd=$0.92\pm0.02$, while only 20\,\% of the background events at this
energy survive.

For the semi-coaxial detectors, a PSD method based on an artificial neural
network (ANN)~\cite{gerda:2013:psd} is used.  The signal acceptance
\effpsd=$0.90^{+0.05}_{-0.09}$ is adjusted with DEP events and the uncertainty
is derived from the \twonu\ spectrum and from events at the Compton
edge. About 55\,\% of the background events around \qbb\ are classified as
SSE-like and considered for the analysis.  Two alternative PSD methods were
developed based on a likelihood ratio and on a combination of $A/E$ and the
asymmetry of the current pulse; they are used for cross-checks.  The three PSD
methods use very different training samples and selection criteria but more
than 90\,\% of the events rejected by ANN are also rejected by the two other
algorithms.

The half-life on \onbb\ decay is calculated as
\begin{eqnarray} \label{eq:counts} 
T_{1/2}^{0\nu} & = & \frac{\ln 2 \cdot N_A}{m_{enr}\cdot N^{0\nu}} \,\cdot 
{ \cal E} \cdot \epsilon  \\  
\epsilon  &=&  
    \fgesix \cdot \factvol \cdot \varepsilon_{fep} \cdot 
  \varepsilon_{psd}   \label{eq:eff}  
\end{eqnarray} 
with $N_A$ being Avogadro's constant, \exposure\ the total exposure (detector
mass $\cdot$ live time), and $m_{enr} = 75.6$~g the molar mass of the enriched
material. $N^{0\nu}$ is the observed signal strength or the corresponding
upper limit.  The efficiency $\epsilon$ accounts for the fraction of
\gess\ atoms (\fgesix), the active volume fraction (\factvol), the signal
acceptance by PSD ($\varepsilon_{psd}$), and $\varepsilon_{fep}$. The latter
is the probability that a \onbb\ decay taking place in the active volume of a
detector releases its entire energy in it, contributing to the full energy
peak at $Q_{\beta\beta}$. Energy losses are due to bremsstrahlung photons,
fluorescence X-rays, or electrons escaping the detector active volume.  Monte
Carlo simulations yield $\varepsilon_{fep}=0.92$ (0.90) for semi-coaxial
(BEGe) detectors.

The \gerda\ background model~\cite{gerda:2013:bck} predicts approximately a
flat energy distribution between 1930 and 2190~keV from Compton events of
$\gamma$ rays of $^{208}$Tl and $^{214}$Bi decays, degraded $\alpha$ events,
and $\beta$ rays from $^{42}$K and $^{214}$Bi. The signal region
($2039\pm5$)~keV and the intervals ($2104\pm5$)~keV and ($2119\pm5$)~keV,
which contain known $\gamma$-ray peaks from $^{208}$Tl and $^{214}$Bi,
respectively, are excluded in the background calculation.  The net width of
the window used for the evaluation of the constant background is hence
230~keV.
 
Data are grouped into three subsets with similar characteristics: ($i$) data
from the BEGe detectors form one set, ($ii$) the {\it golden} data set
contains the major part of the data from the semi-coaxial detectors except
($iii$) two short periods with higher background levels when the BEGe
detectors were inserted ({\it silver} data set).

\section{Results}
\begin{figure}[t] 
\includegraphics[width=1.0\columnwidth]{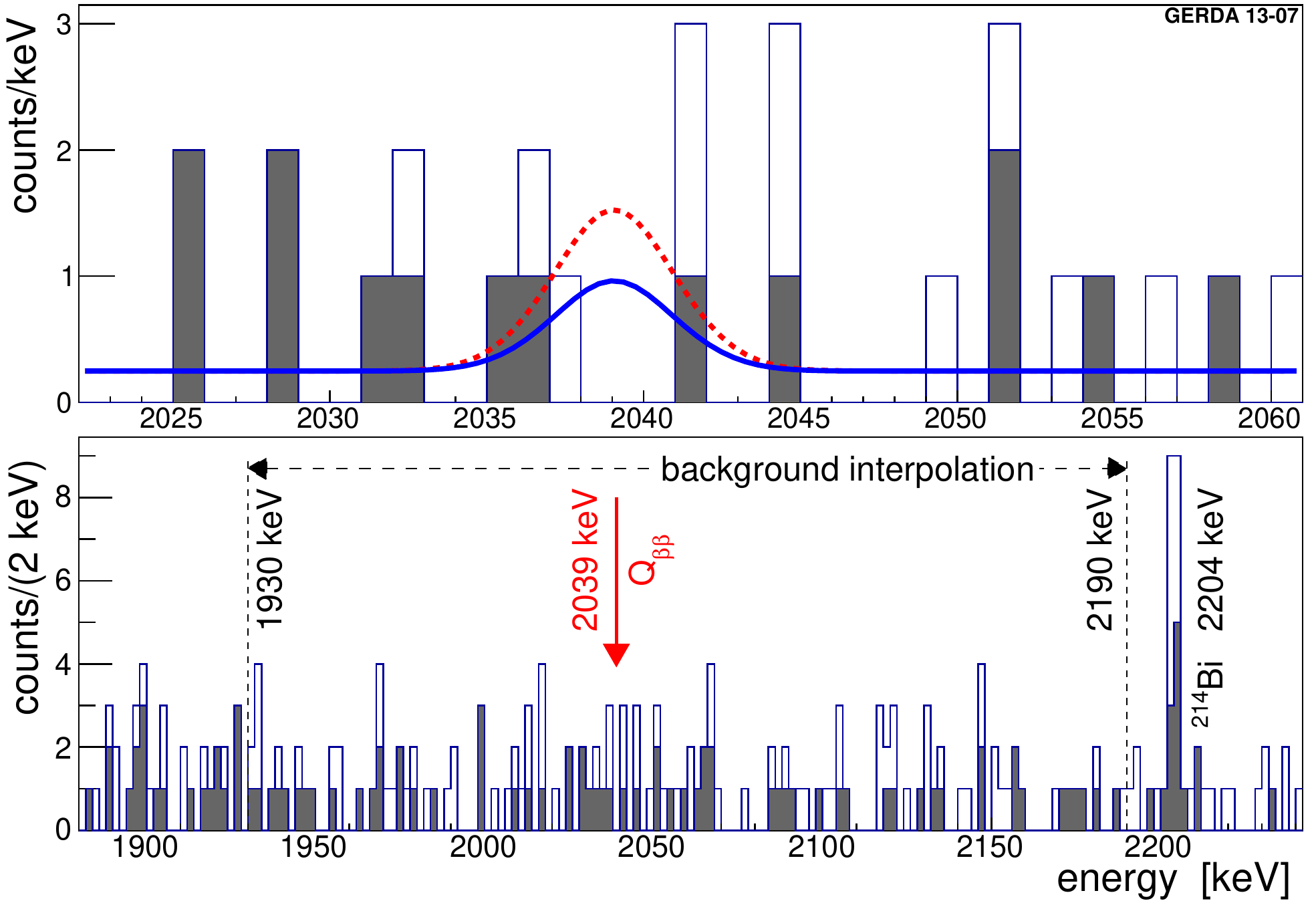} 
\caption{\label{fig:spectrum}
       The combined energy spectrum from all \geenr\ detectors without (with)
       PSD is shown by the open (filled) histogram. The lower panel shows the
       region used for the background interpolation. In the upper panel, the
        spectrum zoomed to \qbb\ is superimposed with the expectations (with
       PSD selection) based on the central value of Ref.~\cite{klapdor1},
       $T_{1/2}^{0\nu} = 1.19\cdot10^{25}$~yr (red dashed) and with the
       90\,\% upper limit derived in this work, corresponding to
       $T_{1/2}^{0\nu} = 2.1\cdot10^{25}$~yr (blue solid).
} 
\end{figure}

Table~\ref{tab:results} lists the observed number of events in the interval
$Q_{\beta\beta}\pm5$~keV for the three data sets, the number of background
events in the 230~keV window and the exposure-weighted average efficiency
$\langle\epsilon\rangle$ over all detectors.  Table~\ref{tab:results2}
reports the details of these events including the results from the PSD
analysis.  The combined energy spectrum around \qbb, with and without the PSD
selection, is displayed in Fig.~\ref{fig:spectrum}.
\begin{table}[t] 
\caption{\label{tab:results} 
    Parameters for the three data sets with and without the pulse shape
    discrimination (PSD). ``bkg'' is the number of events in the 230~keV
    window and BI the respective background index, calculated as
    bkg/(\exposure\ $\cdot$ 230~keV). ``cts''
    is the observed number of events in the interval \qbb$\pm5$~keV.  
} 
\begin{ruledtabular} 
\begin{tabular}{crlrrr} 
data set & \exposure [kg$\cdot$yr] & $\langle\epsilon\rangle$ & bkg 
&BI $^\dagger)$& cts \\  %
\hline 
\multicolumn{6}{l}{without PSD} \\
{\it golden} & 17.9 & $0.688\pm0.031$ & 76 &18$\pm$2 & 5 \\     
{\it silver} & 1.3  & $0.688\pm0.031$ & 19 &63$^{+16}_{-14}$ & 1 \\     
{\it BEGe}   & 2.4  & $0.720\pm0.018$ & 23 &42$^{+10}_{-8}$ & 1 \\     
\hline 
\multicolumn{6}{l}{with PSD} \\
{\it golden} & 17.9 & $0.619^{+0.044}_{-0.070}$ & 45 & 11$\pm$2 &  2 \\
{\it silver} & 1.3  & $0.619^{+0.044}_{-0.070}$ &  9 &30$^{+11}_{-9}$&  1 \\
{\it BEGe}   & 2.4  & $0.663\pm0.022$        & 3 & 5$^{+4}_{-3}$& 0 \\       
\end{tabular}  
\end{ruledtabular} 
$^\dagger)$ in units of \dctsper.\hfill~\hfill  
\end{table}    

Seven events are observed in the range $Q_{\beta\beta}\pm5$~keV before the
PSD, to be compared to $5.1\pm0.5$ expected background counts. No excess of
events beyond the expected background is observed in any of the three data
sets. This interpretation is strengthened by the pulse shape analysis. Of the
six events from the semi-coaxial detectors, three are classified as SSE by
ANN, consistent with the expectation.  Five of the six events have the same
classification by at least one other PSD method. The event in the BEGe data
set is rejected by the A/E cut.  No events remain within
$Q_{\beta\beta}\pm\sigma_E$ after PSD.  All results quoted in the following
are obtained with PSD.

\begin{table}[b] 
\caption{\label{tab:results2} 
    List of all events within $Q_{\beta\beta} \pm 5$~keV
} 
\begin{ruledtabular} 
\begin{tabular}{llrlc} 
data set & detector & energy & date & PSD \\
 & & [keV] & &passed  \\ 
\hline 
{\it golden} & ANG~5 & 2041.8 & 18-Nov-2011 22:52 & no \\
{\it silver} & ANG~5 & 2036.9 & 23-Jun-2012 23:02 & yes \\
{\it golden} & RG~2  & 2041.3 & 16-Dec-2012 00:09 & yes \\
{\it BEGe}   & GD32B & 2036.6 & 28-Dec-2012 09:50 & no \\
{\it golden} & RG~1  & 2035.5 & 29-Jan-2013 03:35 & yes \\
{\it golden} & ANG~3 & 2037.4 & 02-Mar-2013 08:08 & no \\
{\it golden} & RG~1  & 2041.7 & 27-Apr-2013 22:21 & no \\
\end{tabular}  
\end{ruledtabular} 
\end{table}    

To derive the signal strength $N^{0\nu}$ and a frequentist coverage interval,
a profile likelihood fit of the three data sets is performed.  The fitted
function consists of a constant term for the background and a Gaussian peak
for the signal with mean at \qbb\ and standard deviation $\sigma_E$ according
to the expected resolution.  The fit has four free parameters: the backgrounds
of the three data sets and $1/T_{1/2}^{0\nu}$, which relates to the peak
integral by Eq.~\ref{eq:counts}.  The likelihood ratio is only evaluated for
the physically allowed region $T_{1/2}^{0\nu}>0$. It was verified that the
method has always sufficient coverage.  The systematic uncertainties due to
the detector parameters, selection efficiency, energy resolution and energy
scale are folded in with a Monte Carlo approach which takes correlations into
account. The best fit value is $N^{0\nu} = 0$, namely no excess of signal
events above the background.  The limit on the half-life is
\begin{equation}  
T_{1/2}^{0\nu} > 2.1\cdot10^{25}\,\,{\rm yr \,\,\,\,\,\, (90\,\%~~ C.L.)} 
\end{equation}    
including the systematic uncertainty. The limit on the half-life corresponds
to $N^{0\nu} < 3.5$~counts.  The systematic uncertainties weaken the limit by
about 1.5\,\%.  Given the background levels and the efficiencies of
Table~\ref{tab:results}, the median sensitivity for the $90\,\% \,$C.L. limit
is $2.4\cdot10^{25}\,\,{\rm yr}$.

A Bayesian calculation~\cite{bayesDBD} was also performed with the same fit
described above. A flat prior distribution is taken for $1/T_{1/2}^{0\nu}$
between 0 and $10^{-24}$~yr$^{-1}$.  The toolkit BAT~\cite{BAT} is used to
perform the combined analysis on the data sets and to extract the posterior
distribution for \thalfzero\ after marginalization over all nuisance
parameters. The best fit is again $N^{0\nu} = 0$ and the 90\,\% credible
interval is $T_{1/2}^{0\nu} > 1.9\cdot10^{25}$~yr (with folded systematic
uncertainties). The corresponding median sensitivity is $T_{1/2}^{0\nu} >
2.0\cdot10^{25}$~yr.

\section{Discussion}

The \gerda\ data show no indication of a peak at \qbb, i.e.~the claim for the
observation of \onbb\ decay in \gess\ is not supported.  Taking
\thalfzero\ from Ref.~\cite{klapdor1}, $5.9\pm1.4$ decays are expected~(see
note~\footnote{The number of signal counts expected can be evaluated by
  rescaling the number of counts $(28.75\pm6.86)$ from Ref.~\cite{klapdor1}
  for the active exposure (\factvol$\cdot$\exposure). All other efficiency
  factors of Eq.~(\ref{eq:eff}) approximately cancel with the exception of
  \effpsd. The expected number of events after the PSD selection is 6.8 (6.5
  in $Q_{\beta\beta} \pm 2 \sigma_E$). The difference with respect to the
  value calculated from \thalfzero\ is due to the efficiency factor
  $\varepsilon_{fep}$, which is taken to be 100\,\% in Ref.~\cite{klapdor1}.})
in $\Delta E=\pm 2 \sigma_E$ and $2.0\pm0.3$ background events after the PSD
cuts, as shown in Fig.~\ref{fig:spectrum}.  This can be compared with three
events detected, none of them within $Q_{\beta\beta} \pm \sigma_E$.  The model
($H_1$), which includes the claimed \onbb\ signal from Ref.~\cite{klapdor1},
gives in fact a worse fit to the data than the background-only model ($H_0$):
the Bayes factor, namely the ratio of the probabilities of the two models, is
$P(H_1)/P(H_0)=0.024$.  Assuming the model $H_1$, the probability to obtain
$N^{0\nu} = 0$ as the best fit from the profile likelihood analysis is
$P(N^{0\nu}=0|H_1)$=0.01.

The \gerda\ result is consistent with the limits by \hdm\ and \igex.  The
profile likelihood fit is extended to include the energy spectra from
\hdm\ (interval 2000-2080~keV; Fig.~4 of Ref.~\cite{hdm}) and \igex\ (interval
2020-2060~keV; Table~II of Ref.~\cite{igex}).  Constant backgrounds for each
of the five data sets and Gaussian peaks for the signal with common
$1/T_{1/2}^{0\nu}$ are assumed.  Experimental parameters (exposure, energy
resolution, efficiency factors) are obtained from the original references or,
when not available, extrapolated from the values used in \gerda.  The best fit
yields $N^{0\nu} = 0$ and a limit of
\begin{equation}\label{eq:allge}  
T_{1/2}^{0\nu} > 3.0\cdot10^{25}\,\,{\rm yr \,\,\,\,\,\,  (90\,\%~~ C.L.).}  
\end{equation}    
The Bayes factor is $P(H_1)/P(H_0)=2\cdot10^{-4}$; the claim is hence strongly
disfavored.

\begin{figure}[tb] 
\includegraphics[width=1.\columnwidth,angle=270]{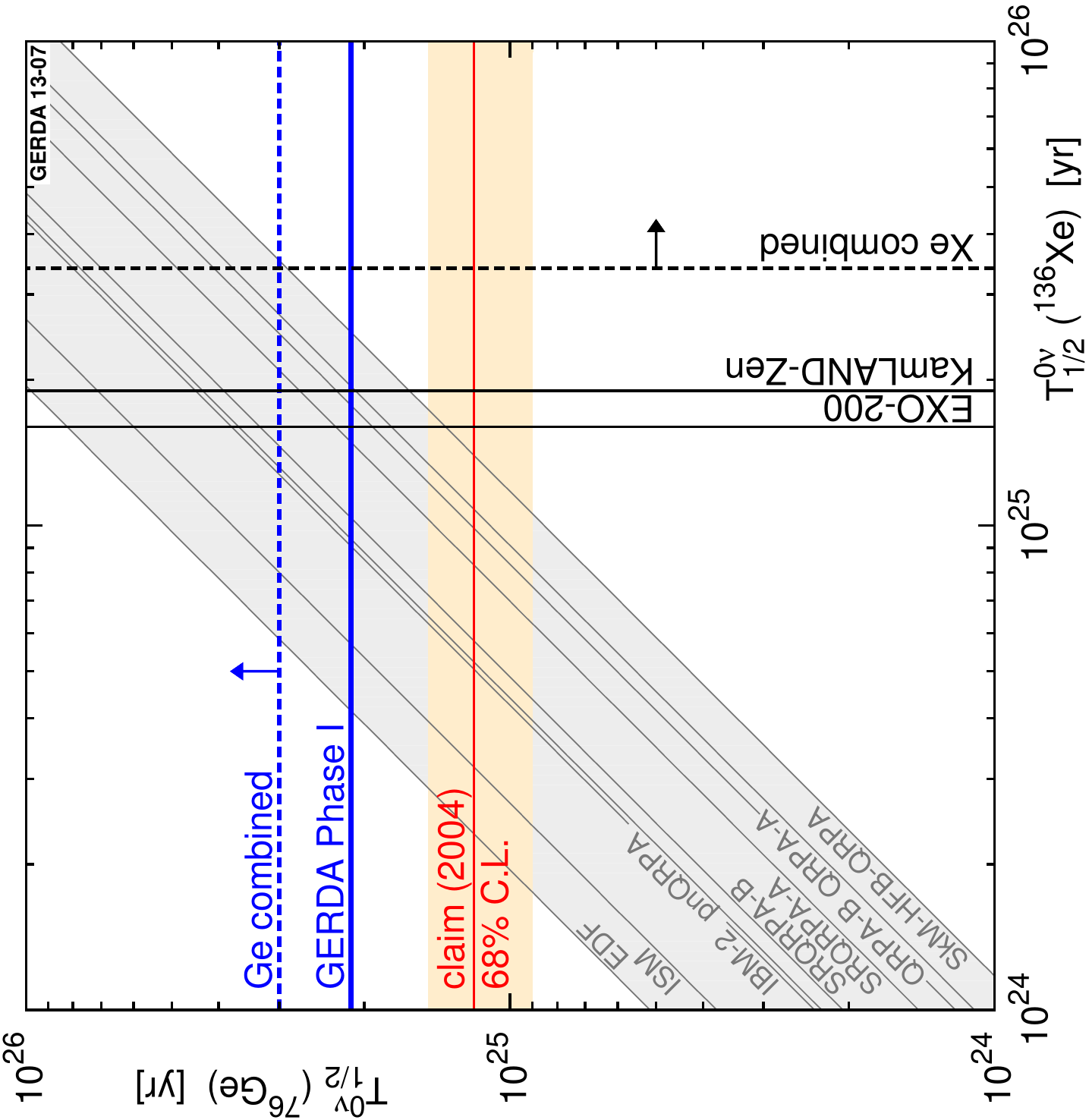}
\caption{\label{fig:comparison}
           Limits (90\,\% C.L.) on \thalfzero\ of \gess\ (this work) and
           $^{136}$Xe~\cite{kamland,exo} compared with the signal claim for
           \gess\ of Ref.~\cite{klapdor1} (68\,\% C.L. band).  The lines in
           the shaded gray band are the predictions for the correlation of the
           half-lives in $^{136}$Xe and in \gess\ according to different NME
           calculations~\cite{rod10,mene09,bar13,suh10,mer13,fae13,mus13}.
           The selection of calculations
            and the labels are taken from Ref.~\cite{bup13}.
}  
\end{figure}

Whereas only \gess\ experiments can test the claimed signal in a
model-independent way, NME calculations can be used to compare the present
\gess\ result to the recent limits on the $^{136}$Xe half-life from
KamLAND-Zen~\cite{kamland} and EXO-200~\cite{exo}.  Fig.~\ref{fig:comparison}
shows the experimental results, the claimed signal (labeled ``claim (2004)'')
and the correlations for different predictions, assuming that the exchange of
light Majorana neutrinos is the leading mechanism. Within this assumption, the
present result can be also combined with the $^{136}$Xe experiments to
scrutinize Ref.~\cite{klapdor1}. The most conservative exclusion is obtained
by taking the smallest ratio
$M_{0\nu}$($^{136}$Xe)/$M_{0\nu}$($^{76}$Ge)$\simeq 0.4$~\cite{fae13,mus13} of
the calculations listed in Ref.~\cite{bup13}.  This leads to an expected
signal count of 23.6$\pm$5.6 (3.6$\pm$0.9) for KamLAND-Zen (EXO-200). The
comparison with the corresponding background-only models~\footnote{The
  sensitivity of KamLAND-Zen corresponds to an equivalent background of
  460$\pm$21.5 counts. The equivalent observed counts are -1.17$\sigma$ lower,
  i.e.~about 435 events~\cite{berg13}.  EXO-200 expects $7.5\pm0.7$ counts in
  the interval $Q_{\beta\beta} \pm 2\sigma_E$ and observes 5 events.} yields a
Bayes factor $P(H_1)/P(H_0)$ of 0.40 for KamLAND-Zen and 0.23 for EXO-200.
Including the \GERDA\ result, the Bayes factor becomes 0.0022.  Also in this
case the claim is strongly excluded; for a larger ratio of NMEs the exclusion
becomes even stronger.  Note, however, that other theoretical approximations
might lead to even smaller ratios and thus weaker exclusions.

The range for the upper limit on the effective electron neutrino mass
$m_{\beta\beta}$ is 0.2 - 0.4~eV.  This limit is obtained by using the
combined \gess\ limit of Eq.~\ref{eq:allge}, the recently re-evaluated phase
space factors of Ref.~\cite{kot12} and the NME calculations mentioned
above~\cite{rod10,mene09,bar13,suh10,mer13,fae13,mus13}. Scaling due to
different parameters $g_A$ and $r_A$ for NME is obeyed as discussed in
Ref.~\cite{anatoly}.

In conclusion, due to the unprecedented low background counting rate and the
good energy resolution intrinsic to HPGe detectors, \GERDA\ establishes after
only 21.6~\kgyr\ exposure the most stringent \onbb\ half-life limit for \gess.
The long-standing claim for a \onbb\ signal in \gess\ is strongly disfavored,
which calls for a further exploration of the degenerate Majorana mass
scale. This will be pursued by \gerda\ Phase~II aiming for a sensitivity
increased by a factor of about 10.

\appendix
\section{Acknowledgments}       
 The \gerda\ experiment is supported financially by
   the German Federal Ministry for Education and Research (BMBF),
   the German Research Foundation (DFG) via the Excellence Cluster Universe,
   the Italian Istituto Nazionale di Fisica Nucleare (INFN),
   the Max Planck Society (MPG),
   the Polish National Science Centre (NCN),
   the Foundation for Polish Science (MPD programme),
   the Russian Foundation for Basic Research (RFBR), and
   the Swiss National Science Foundation (SNF).
 The institutions acknowledge also internal financial support.

The \gerda\ collaboration thanks the directors and the staff of the LNGS
for their continuous strong support of the \gerda\ experiment.

\providecommand{\noopsort}[1]{}\providecommand{\singleletter}[1]{#1}%

\end{document}